\documentclass[reprint,amsmath,amssymb]{revtex4-1}
\usepackage[utf8x]{inputenc}
\usepackage{graphicx}% Include figure files
\usepackage{dcolumn}% Align table columns on decimal point
\usepackage{bm}% bold math
\usepackage{color} 
\usepackage{amsmath}% bold math
\usepackage{tensor}% bold mathpp

\usepackage{soul} % strikethrough
\usepackage[normalem]{ulem} % strikethrough

\begin{document}

\preprint{APS/123-QED}

%\title{Localization in quasi-periodic systems from calculations with periodic boundary conditions}

\title{Scaling of the bulk polarization in extended and localized phases of a quasiperiodic model}

\author{Bal\'azs Het\'enyi}
\affiliation{MTA-BME Quantum Dynamics and Correlations Research Group, Department of Physics, Budapest University of Technology and
  Economics, H-1111 Budapest, Hungary \\ and \\
 Institute for Solid State Physics and Optics, Wigner Research Centre for Physics,  H-1525 Budapest, P. O. Box 49, Hungary}

\date{\today}% It is always \today, today,

\begin{abstract}
\textcolor{black}{We study the finite size scaling of the bulk polarization in a quasiperiodic (Aubry-Andr\'{e}) model using the geometric analog of the Binder cumulant.   As a proof of concept we show that the geometric Binder cumulant method described here can reproduce the known literature values for the flat and raised cosine distributions, which are the two distributions that occur in the delocalized phase.   For the Aubry-Andr\'{e} model at half-filling the phase transition point is accurately reproduced.  Not only is the correct size scaling exponent of the variance obtained in the extended and the localized phases, but the geometric Binder cumulant undergoes a sign change at the phase transition.  We also calculate the state resolved Binder cumulant as a function of disorder strength to gain insight into the mechanism of the localization transition.}
\end{abstract}


\maketitle

\section{Introduction}

The modern polarization theory~\cite{King-Smith93,Resta94,Vanderbilt18,Resta00}  (MPT) overcomes the problem of the ill-defined nature of the position operator in systems with periodic boundary conditions (PBC) by casting crystalline polarization as a  geometric or Berry~\cite{Berry84} phase.  In band insulators, polarization is an open path~\cite{Zak89} geometric phase (also known as a Zak phase), while in many-body systems~\cite{Resta98} it is a single-point geometric phase~\cite{Resta00}, the phase of the expectation value of a unitary operator.  It is also possible to access moments and cumulants of the polarization~\cite{Resta99,Souza00}.  The Zak phase can be viewed as the first member of a cumulant series.  Gauge invariant cumulants~\cite{Souza00} characterize band insulators, while the variance of the polarization, as derived~\cite{Resta99} by Resta and Sorella (RS), provides a criterion for distinguishing~\cite{Kohn64} localized systems from extended ones.  MPT is also the starting point~\cite{Bernevig13,Asboth16} for deriving the characteristic invariants of topological insulators.  Moments of the polarization form the backbone of maximally localized Wannier functions~\cite{Marzari97,Marzari12}.\\

In the study of phase transitions finite size scaling~\cite{Cardy96,Fisher72a,Fisher72b,Binder81a,Binder81b} is a useful tool.  The finite systems accessible in calculations provide the expectation values of physical quantities (for example the magnetization in the Ising model), as well as their moments and cumulants, but care must be taken when extrapolating to the thermodynamic limit.  Using the finite size scaling hypothesis~\cite{Fisher72a,Fisher72b}, one can calculate the Binder cumulant~\cite{Binder81a,Binder81b} (BC).   BC is a ratio of statistical cumulants of the order parameter (represented in quantum systems by an operator) useful in locating critical points.   BC is equivalent to the {\it excess kurtosis}~\cite{Shynk13} (EK) of the order parameter, a quantity widely  used in probability and statistics to characterize the tail and peak of probability distributions.  An EK of zero corresponds to a Gaussian distribution, while a distribution with positive(negative) EK is referred to as  super-Gaussian(sub-Gaussian).  In MPT  the polarization is not an expectation value of an operator, it is  a geometric phase, so it is not immediately obvious whether a BC can be constructed for the polarization in crystalline systems.   \\

 MPT, in its original form, does not provide accurate finite size scaling information in the delocalized phase.   For disordered systems~\cite{Abrahams79,Langedijk09} or in many-body localization studies~\cite{Basko06,Nandkishore15} the inverse participation ratio is used instead of MPT, even though MPT was developed to address the localization of charge carriers.   Kerala Varma and Pilati~\cite{Varma15} made a thorough investigation of  RS in quasiperiodic models.   They find some  disagreement in the behavior of the variance between open boundary conditions (OBC) and PBC.   RS can be interpreted as a second cumulant (variance) obtained from a particular generating function.  It is possible to extend to RS formalism to derive higher order cumulants as well~\cite{Hetenyi19,Hetenyi22} but these diverge in metallic phases and it is not possible to construct a BC.  Recently an alternative approximation scheme~\cite{Hetenyi19,Hetenyi22}, which preserves the scaling information of the variance in the metallic, as well as the insulating phases, was suggested to circumvent these problems.  Since the approach amounts to formulating a Binder cumulant in the context of a geometric phase, the method will be named the geometric Binder cumulant (GBC).\\
 
\begin{figure}[ht]
 \centering
 \includegraphics[width=8cm,keepaspectratio=true]{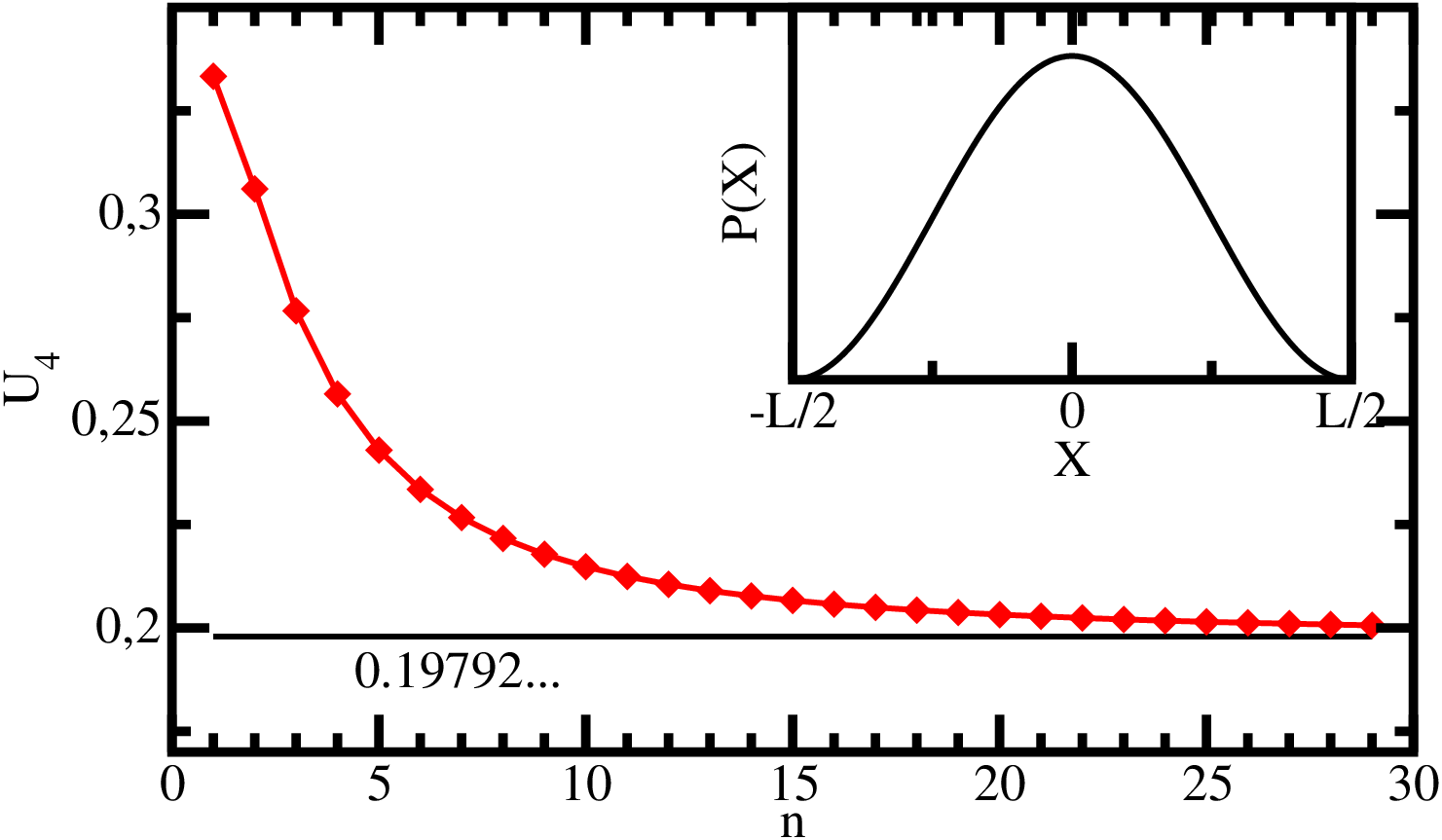}
 \caption{Binder cumulant for a closed gap system with a two-fold degenerate ground state as a function of $n$.  The value of  $n$ determines the order of the finite difference approximation ($\mathcal{O}(L^{-2n})$).  As the approximation is improved, the geometric Binder cumulant approaches the limiting value of $0.19792...$ which corresponds to the known~\cite{Rinne10} value of the excess kurtosis for the raised cosine distribution.}
 \label{fig:U4raisedcos}
\end{figure}

 \textcolor{black}{In this work we apply the GBC to the Aubry-André~\cite{Aubry80} (AA) model.  The AA model is used to study quasicrystals~\cite{Martinez18,Billy08,Roati08} and the localization phenomenon~\cite{Dominguez-Castro19,Jitomirskaya99,Modugno09,Wang17,Zhang15}.  It exhibits a phase transition between extended and localized states at a finite on-site potential potential strength.   At finite system size we find that the ground state on the metallic side can be degenerate.   We discuss ways to handle this issue.  The distribution function of the polarization in the degenerate case is a raised cosine distribution (RCD), whose EK is a specific number known from the statistics literature~\cite{Rinne10}.  Our lowest order approximation does not reproduce that known value, but as higher order finite difference approximations are used the fourth order GBC ($U_4$) converges to it (Fig. \ref{fig:U4raisedcos}).  The GBC also characterizes the critical exponents near the AA transition and locates the transition point accurately (Figs. \ref{fig:M2} and \ref{fig:U4}). } \\
 
 \textcolor{black}{We also calculate the GBC for individual eigenstates (Fig. \ref{fig:U4_state}) and find that all states are delocalized in the extended phase, while in the localized phase, only {\it some} states become localized, but this causes the entire system (at half filling) to localize at finite fillings.  For the AA model Jitomirskaya~\cite{Jitomirskaya99} has shown that for infinite systems all states localize at the transition point ($W=2t$).  Recently this result was questioned to some extent by the suggestion~\cite{Wang17,Zhang15} of "almost localized" states.  Our results within periodic boundary conditions suggest that some states remain delocalized even for $W>2t$, but this may be due to the periodic boundary conditions.  It is to be noted that a calculation of this type is not possible without the use of the technique developed here: our technique allows the determination of whether a state is localized or not from one calculation on the state itself, without having to compare different system sizes. } \\

Our paper is organized as follows.  In the next section the MPT based cumulant method is assembled.  In section \ref{sec:model} the model and the specificities of the calculation are chronicled.  We then relate a Widom scaling analysis, in which we derive and then numerically verify a universal relation between critical exponents.  In section \ref{sec:deg} we address the effects of ground state degeneracy.  In section \ref{sec:results} the results of our calculations are presented.  In section \ref{sec:conclusion} we conclude our work.

\section{The geometric Binder cumulant}
\label{sec:MC}

\textcolor{black}{In this section we construct a version of the Binder cumulants adapted to  the context of the modern polarization theory (MPT), which we will refer to as the geometric Binder cumulant (GBC).   We briefly introduce the MPT by way of the twist operator, and then focus on how cumulants are constructed.  Moments and cumulants are derivatives of the generating function, but in a periodic system this function is only defined on a discrete set of $k$-points, so one can not take an exact derivative.  Given that, the usual approach is to take finite difference derivatives, and we argue that it is crucial how the finite difference approximation is applied, because not all approximation schemes can reproduce finite size scaling information.  Our approach is validated through the reproduction of  the known value of the excess kurtosis for a number of distribution functions (Fig. 1 of Ref. \cite{Hetenyi22} and Fig. \ref{fig:U4raisedcos} of this work) found in the delocalized phase.  As for the insulating phase, it was shown~\cite{Hetenyi22} that the GBC ($U_4$) can only take nonzero values for adiabatic paths which cross degeneracy points, and it is always zero for fully adiabatic paths.  Based on this one can expect that in the insulating phase the GBC always takes a value of zero.  Our numerical results show that this happens in the limit of large system sizes.  A GBC of zero corresponds to a Gaussian distribution of the polarization.}

\subsection{The twist operator}

In MPT the position operator is not used directly.  Instead, for the electronic contribution~\cite{note} of the polarization, expectation values of the twist operator,
\begin{equation}
\label{eqn:U}
\hat{U} = \exp \left( i \frac{2 \pi}{L} \hat{X} \right),
\end{equation}
are taken, from which the total position and its cumulants can be extracted.  In Eq. (\ref{eqn:U}) $L$ is the length of the system in which it is periodic, $\hat{X} = \sum_{j=1}^N x_j \hat{n}_j$ is the total position operator ($x_j$ denotes the position, $\hat{n}_j$ denotes the density operator at site $j$, and $N$ is the number of particles).  The expectation value over some ground state $\Psi$,
\begin{equation}
Z_q = \langle \Psi | \hat{U}^q | \Psi \rangle
\end{equation}
can be interpreted as a characteristic (cumulant generating) function, associated with the probability distribution of the total position,
\begin{equation}
P(X) = \langle \Psi |\delta(X - \hat{X})| \Psi \rangle.
\end{equation}
Since $P(X)$ is periodic in $L$, $Z_q$ is only defined on a discrete set of points $q=0,...,L-1$.    In general, $\Psi$ denotes a correlated ground state.  In band systems, such as the Aubry-Andr\'{e} model we study, $\Psi$ is a Slater determinant constructed from occupied Bloch states. 

\subsection{Cumulants in statistics}

Given a normalized probability distribution function, $P_0(x)$, which satisfies
\begin{equation}
\label{eqn:P0}
P_0(x) \geq 0, \int_{-\infty}^\infty dx P_0(x) = 1,
\end{equation} 
one can define the associated characteristic function,
\begin{equation}
f(k) = \int_{-\infty}^\infty d x \exp(i k x) P_0(x).
\end{equation}
The $n$th moment ($M_n$) and the $n$th cumulant ($C_n$) of $P(x)$ can be obtained from $f(k)$ as
\begin{eqnarray}
M_n &=& \frac{1}{i^n} \left.\frac{\partial^n f(k)}{\partial k^n}\right|_{k=0} = \langle x^n \rangle.
 \\ \nonumber
 C_n &=& \frac{1}{i^n} \left.\frac{\partial^n \ln f(k)}{\partial k^n}\right|_{k=0},
 \end{eqnarray}
 where $\langle \rangle$ denote the average over $P_0(x)$.  The moments and cumulants are related to each other, the first few such relations can be written,
 \begin{eqnarray}
 \label{eqn:C_M}
 C_1 &=& M_1 \\ \nonumber
 C_2 &=& M_2 - M_1^2 \\ \nonumber
 C_3 &=& M_3 - 3 M_2 M_1^2 + 2 M_1^3 \\ \nonumber
 C_4 &=& M_4 - 4 M_3 M_1 - 3 M_2^2 + 12 M_2 M_1^2 - 6 M_1^4.
 \end{eqnarray}
 $C_1$ is known as the mean, $C_2$ is the variance, $C_3$ is the skew, and $C_4$ is the kurtosis.   The distribution, $P(x)$ can be shifted so that the mean is zero.  In this case the cumulants and moments become centered, and the relations in Eq. (\ref{eqn:C_M}) become,
 \begin{eqnarray}
 \label{eqn:C_M}
 C_1 &=& 0 \\ \nonumber
 C_2 &=& M_2  \\ \nonumber
 C_3 &=& M_3  \\ \nonumber
 C_4 &=& M_4 - 3 M_2^2.
 \end{eqnarray}
 The excess kurtosis, used in statistics to characterize the tails of distribution functions, is defined as,
 \begin{equation}
 K_E = \frac{C_4}{C_2^2}.
 \end{equation}
 When the quantity $x$ refers to the order parameter of a physical system, the fourth order Binder cumulant ($U_4$) is often used in finite size scaling,
 \begin{equation}
 U_4 = 1 - \frac{M_4}{3 M_2^2} = -\frac{1}{3} K_E,
 \end{equation}
 which is equivalent to the excess kurtosis.  The two quantities only differ in the factor of $-\frac{1}{3}$.\\
 
 \subsection{The case of periodic probability distribution functions}
 
 In the modern theory of polarization, crystalline systems are treated, whose Hamiltonians are taken as periodic.  The underlying probability distributions are also periodic.   We can write
 \begin{equation}
 P(x) = \sum_{w=-\infty}^\infty P_0(x+wL),
 \end{equation}
 where $P_0(x)$ is a normalized probability distribution, not periodic, as defined in Eq. (\ref{eqn:P0}).  $P(x)$ is periodic in $L$.  One is interested in the moments and cumulants of $P_0(x)$, but the characteristic function is only available at a discrete set of $k$ points,
 \begin{equation}
 k_q = \frac{2 \pi}{L}q,
 \end{equation}
where $q = 0,...,L-1$.  This also means that the derivatives which need to be calculated to obtain $C_n$ or $M_n$ have to be approximate derivatives.  Most often they are finite difference derivatives.  Denoting the discrete characteristic function as $Z_q$, which we define as,
\begin{equation}
\label{eqn:Zq_FT}
Z_q = \int d x P_0(x) \exp\left( i \frac{2 \pi }{L}q x \right).
\end{equation}
The Resta-Sorella approximation for the variance amounts to using a finite difference approximation for $C_2$ of order $\mathcal{O}(L^{-2})$,
\begin{equation}
\label{eqn:C2_2}
C_2 = \left(\frac{L}{2 \pi i}\right)^2 \left( \ln Z_1 + \ln Z_{-1} - 2 \ln Z_0\right) = -\frac{L^2}{2 \pi^2} \ln |Z_1|.
\end{equation}
In the last part of this equation we used the fact that $Z_{-1} = Z_1^*$, and that $Z_0=1$.

\subsection{Finite difference derivatives}

\textcolor{black}{Given a function $f(x)$ represented on a discrete set of points, $f_q = f(qh)$, with $q$ integer, where  $h$ denotes the spacing between the points.  One can construct~\cite{Hildebrand68} the centered finite difference approximation of $f_q$ to different levels of accuracy.  For example, for the second derivative at $q=0$, we can write 
\begin{equation}
\left. \frac{\partial^2 f(x)}{\partial x^2} \right|_{x=0}\approx \frac{f_1 + f_{-1} - 2 f_0}{h^2}.
\end{equation}
This approximation leads to an error of order $\mathcal{O}(h^{-2})$.  The Resta-Sorella variance can be derived using this approximation, by associating $h = \frac{2 \pi}{L}$ and $\ln |Z_q| = f_q$.  One can also improve on the Resta-Sorella scheme, by using higher order approximations to the second derivative, for example, the next order ($\mathcal{O}(h^{-4})$) is,
\begin{equation}
\left. \frac{\partial^2 f(x)}{\partial x^2} \right|_{x=0}\approx\frac{-f_2 + 16 f_1 - 30 f_0 + 16 f_{-1} - f_{-2}}{12 h^2}.
\end{equation}
We see that in higher order approximations $f_q$s with larger $q$s appear.   Using this approximation one can obtain an $\mathcal{O}(h^{-4})$ approximation to the variance,
\begin{equation}
C_2 = \frac{L^2}{24 \pi^2} (\ln |Z_2| - 16\ln |Z_1|).
\end{equation}
By including more $Z_q$s with larger $q$s one can obtain systematically improved approximations in the insulating phase.}\\

\textcolor{black}{It is also possible to derive higher order cumulants, such as the kurtosis, needed to construct the Binder cumulant.  The lowest order ($\mathcal{O}(h^{-2})$) approximation to the fourth derivative reads as,
\begin{equation}
\left. \frac{\partial^4 f(x)}{\partial x^4} \right|_{x=0}\approx \frac{f_2 -4 f_1 + 6 f_0 -4 f_{-1} + f_{-2}}{h^4}.
\end{equation}
Based on this approximation one can write an expression for the kurtosis of the polarization as
\begin{equation}
\label{eqn:C4_2}
C_4 = \frac{L^4}{8 \pi^4}(\ln |Z_2| - 4 \ln |Z_1|),
\end{equation}
alternatively, using the next best approximation results in
\begin{equation}
C_4 =  \frac{L^4}{48 \pi^4}(-\ln|Z_3| +12 \ln |Z_2| - 39 \ln |Z_1|).
\end{equation}
Again, higher order approximations need $Z_q$s with progressively larger values of $q$.  In principle, one is in a position now to construct the Binder cumulant, since both $C_4$ and $C_2$ are available.  However, if one used the above approximations, one would encounter a problem, namely that the terms $Z_q$ approach zero in the extended phase, leading to divergences in $\ln Z_q$.} \\

\subsection{Approximating the logarithm}

\textcolor{black}{It is, of course, possible to approximate the terms $\ln |Z_q|$, by expanding the quantity $(|Z_q|-1)$, resulting in
\begin{equation}
\label{eqn:ln_approx}
\ln [1 + (|Z_q| - 1)] = (|Z_q|-1) - \frac{(|Z_q|-1)^2}{2} + ...
\end{equation}
Using this approximation in $\mathcal{O}(L^{-2})$ expressions for $C_4$ (Eq. (\ref{eqn:C4_2})) and $C_2$ (Eq. (\ref{eqn:C2_2})) lead to 
\begin{eqnarray}
\label{eqn:C2C4}
C_2 &=& \frac{L^2}{2 \pi^2} (1 - |Z_1|)\\
C_4 &=& \frac{L^4}{8 \pi^4} \left( - |Z_2| + 4 |Z_1| - 3\right). \nonumber
\end{eqnarray}
We define the geometric Binder cumulant (GBC) as
\begin{equation}
\label{eqn:U4}
U_4 = -\frac{1}{3} \frac{C_4}{C_2^2}.
\end{equation}
In the nondegenerate case of the extended phase all $Z_q \rightarrow 0$, except $Z_0$ which is unity.  It follows that $U_4 = \frac{1}{2}$.  In the degenerate case, as discussed below, $Z_q\rightarrow 0$, except for $Z_0 = 1$ and $Z_1 = 1/2$, leading to $U_4=\frac{1}{3}$.   In the extended state $\ln Z_q$ for $q>1$ will diverge.  $C_2$ will not diverge at $\mathcal{O}(L^{-2})$, but if higher order approximations are used, it will.  Also $C_4$ will diverge, and so will $U_4$.  This means that in the original RS scheme size scaling information in the extended phase will be lost.  Approximating $\ln Z_q$ as in Eq. (\ref{eqn:ln_approx}) solves this problem. } \\

\textcolor{black}{It may appear that the GBC technique advocated here is worse than the original RS scheme, since another approximation is introduced.  It is to be noted that the two simple polarization distributions which occur in the extended phase (the flat distribution in the non-degenerate case, and the RCD in the degenerate case) both have known EK values, which are reproduced by the GBC technique (Fig. 1 in Ref. \cite{Hetenyi22} and Fig. \ref{fig:U4raisedcos}), but this can not done by the original RS scheme, which exhibits divergences. }\\

\section{Model and calculation details}
\label{sec:model}

We consider a Hamiltonian of the Aubry-Andr\'{e} type given by
\begin{equation}
\label{eqn:HAA}
\hat{H} = -t \sum_{j=1}^L (c_{j+1}^\dagger c_j + c_j^\dagger c_{j+1}) + W \sum_{j=1}^L \xi_j n_j,  
\end{equation}
where $\xi_j = \cos \left( 2 \pi \alpha j \right)$, where $\alpha$ is the golden ratio obtained from the ratio of consecutive members of the Fibonacci sequence in the limiting case, and the operators $c_j^\dagger$($c_j$) create(annihilate) a particle at site $j$.   For finite systems with PBC, the irrational $\alpha$ is approximated as a ratio $\alpha \approx F_{n+1}/F_n$, where $F_{n}$ is the $n$th Fibonacci number, so the size of the system can not be smaller than $F_n$. \\ 

\textcolor{black}{In our calculations below, we diagonalize $\hat{H}$ under periodic boundary conditions, meaning that we obtain a set of states on the lattice, $\Phi_\lambda(j)$, where $\lambda$ denotes the state index, and $j$ denotes the lattice site.  For a system with $N$ particles, the ground state wave function is a Slater determinant consisting of the $N$ states, $\Phi_\lambda(j)$, $\lambda = 1,..,N$ with lowest energy,
\begin{equation}
\Psi(j_1,...,j_N) = \mbox{Det} \left[ \Phi_\lambda(j_\mu) \right],
\end{equation}
where $j_\mu$ denotes the lattice coordinate of particle $\mu$, and $\mu = 1,...,N$.
To calculate the quantity $Z_q$, one can use the fact that the overlap of determinants equals the determinant of overlaps, resulting in,
\begin{equation}
Z_q = \langle \Psi | \hat{U}^q | \Psi \rangle =  \mbox{Det} \left[ \langle \Phi_\lambda |\hat{u}^q| \Phi_{\lambda'} \rangle \right].
\end{equation},
where $\hat{u}$ denotes the one-body analog of $\hat{U}$ resulting in,
\begin{equation}
 \langle \Phi_\lambda |\hat{U}^q| \Phi_{\lambda'} \rangle = \sum_{j=1}^L \phi^*_\lambda(j) \exp\left( i \frac{2 \pi q}{L}j\right) \phi_{\lambda'}(j).
\end{equation}}

\begin{figure}[ht]
 \centering
 \includegraphics[width=8cm,keepaspectratio=true]{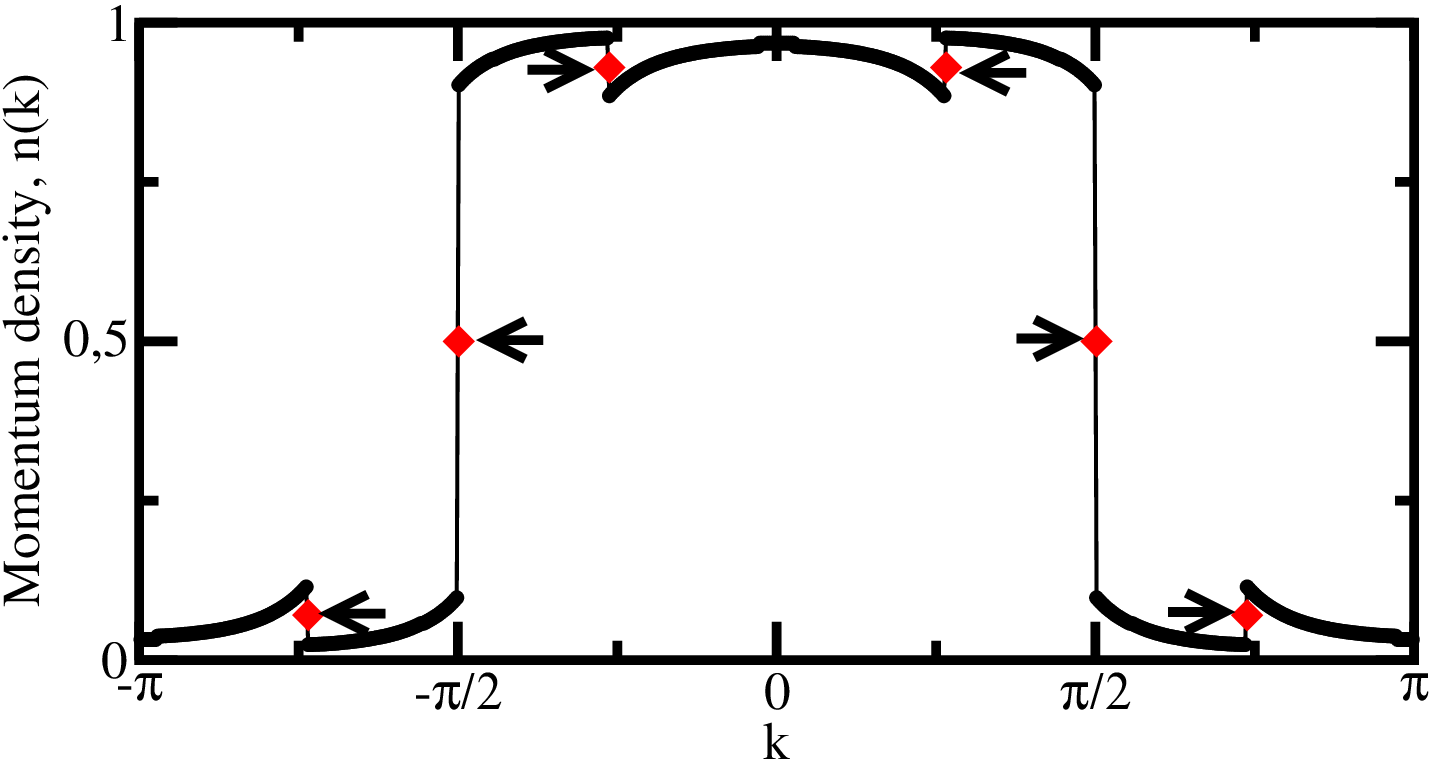}
 \caption{Momentum density~\cite{Fazekas99} as a function of $k$ across the Brillouin zone for an Aubry-Andr\'{e} model with 610 lattice sites for $t=W=1$ (extended phase).   The momentum density has discontinuities as a function of $k$, at $k = \pm \frac{\pi}{4}, \pm \frac{\pi}{2}, \pm \frac{3 \pi}{4}$.  Filled black circles indicate a system with Peierls phase $\phi = 0$, while the red diamonds indicate the states that appear when a Peierls phase of $\phi = \pi/L$ is applied to shift the momenta of states.  These extra states, which always appear halfway through discontinuities are indicated by six arrows. }
 \label{fig:nk}
\end{figure}

\begin{figure}[ht]
 \centering
 \includegraphics[width=8cm,keepaspectratio=true]{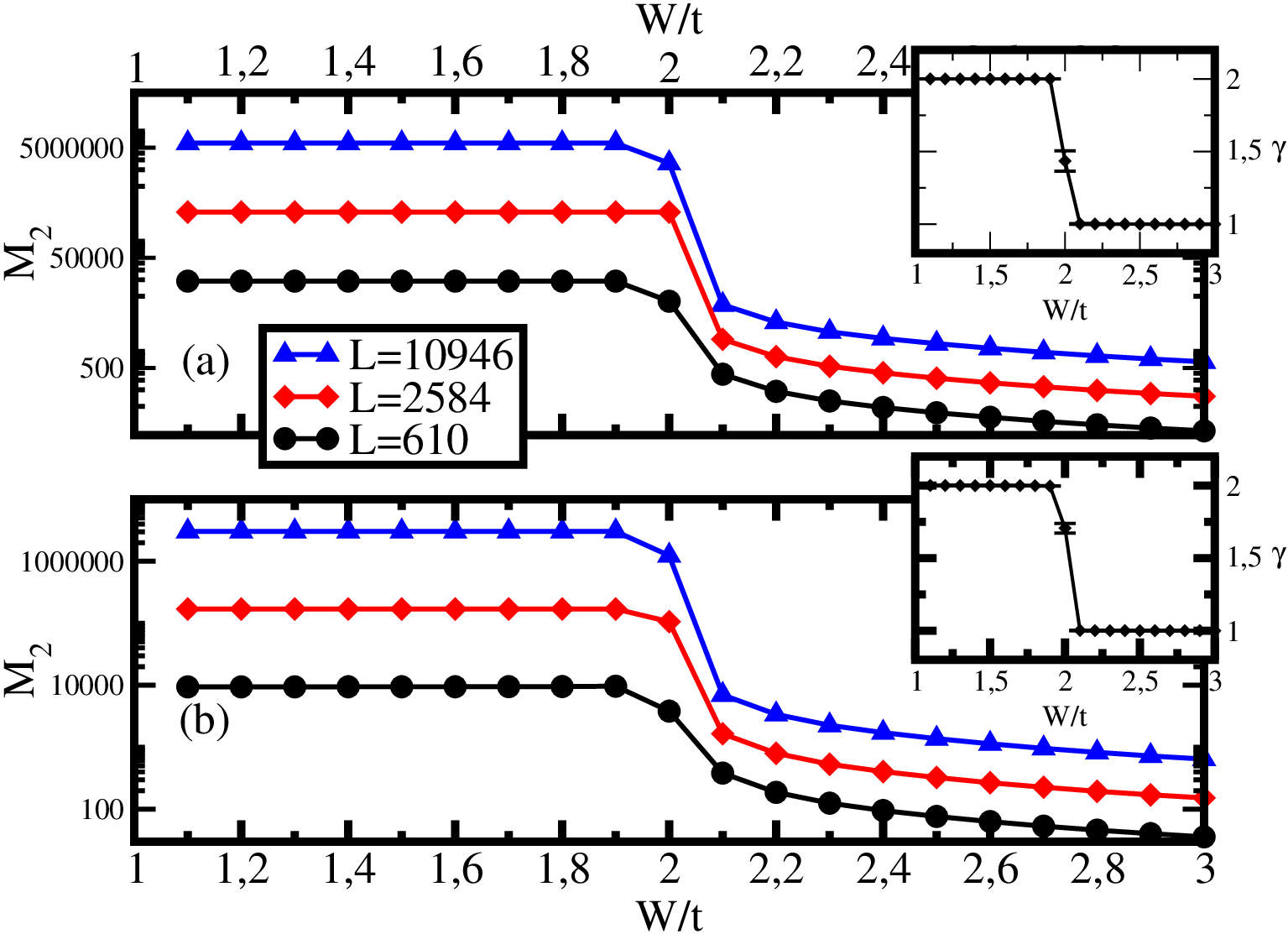}
 \caption{(a) $M_2$ (second cumulant) as a function of $W/t$ for system sizes $L=610, 2584, 10946$.  For $L=610,10946$ $\phi=0$ (corresponding to periodic boundary conditions), while for $L=2584$, $\phi=\pi/L$ (corresponding to anti-periodic boundary conditions).  The ground state in all three cases is non-degenerate.  (b) same as (a) except the Peierls phase is $\phi=\pi/L$ for $L=610, 10946$, and $\phi=0$ for $L=2584$.  The ground state in all three cases is degenerate.  }
 \label{fig:M2}
\end{figure}

\section{Widom scaling in the Aubry-Andr\'{e} model}  
\label{sec:widom}

By considering the continuous generalization of $Z_q$, which occurs in the thermodynamic limit, it is possible to construct a scaling theory within the MPT.   This was done in Ref. \cite{Hetenyi21}.  Sending $\frac{2\pi}{L} q \rightarrow K$, we define the singular "free energy" $\Phi(w,K) = \ln Z(w,K)$, where $w = \frac{W-W_c}{t}$ denotes the reduced potential strength in the vicinity of the transition ($W_c=2$).  The second and fourth cumulants (susceptibilities) all diverge as $W_c$ is approached.   Using the usual definition for susceptibilities, 
\begin{equation}
\chi^{(n)} = \frac{1}{i^n} \frac{\partial^n \Phi(w,K)}{\partial K^n},
\end{equation}
one can define critical exponents which characterize the system in the vicinity of $W_c$, 
\begin{eqnarray}
\chi^{(2)}(w,0) &\propto& 1/w^{\beta}, \\ 
 \chi^{(4)}(w,0) &\propto& 1/w^{\alpha}, \nonumber \\
 \chi^{(2)}(0,K) &\propto& 1/K^{\delta}. \nonumber
 \end{eqnarray}   
Assuming the Widom scaling form, $\Phi(\lambda^a w, \lambda^b K) = \lambda \Phi(w,K)$ leads to the relation between the critical exponents,
\begin{equation}
\label{eqn:abd}
\alpha \delta = \beta (\delta + 2).
\end{equation}
In Ref. \cite{Hetenyi21} the exponent $\delta$ was determined to be $\delta=2$, since this exponent characterizes the system at the critical point, and the distribution of the extended state can be used to estimate it for any model with a localization transition.  This results in the relation $\alpha = 2 \beta$, which appears to be a universal relation in 1D systems for transitions accompanied by gap closure.   For other 1D models this relation was verified in Ref. \cite{Hetenyi21}.  Here, we also find verification for the AA model: we find $\alpha = 2$ and $\beta = 1$, in agreement with Eq. (\ref{eqn:abd}).  \\

\section{Ground state degeneracy}   
\label{sec:deg}

\textcolor{black}{In 1D lattice models a degeneracy often occurs for finite systems, see for example, Ref. \cite{Yang66}.  The origin of this degeneracy is the fact that the Brillouin zone for a finite system is discrete, and whether it occurs or not depends on the parity of $N$ and/or $L$.  This effect is expected to disappear in the thermodynamic limit, and is therefore considered an artifact.  Still, calculations are done in finite systems, so it is important to understand the effects of this degeneracy.  In this section, we discuss how it effects the GBC, and also present various ways of addressing it.  }

In the extended state we find that the GBC can take two values, $U_4 = \frac{1}{2}$ or $U_4 = \frac{1}{3}$.   The former(latter) corresponds to a nondegenerate(degenerate) ground state.  In the nondegenerate case, all $Z_q=0$, except, $Z_0 = 1$.  When the ground state is degenerate, $Z_0=1$, $|Z_1|=1/2$, and all other $Z_q$ are zero.  The degeneracy depends on the optimal spacing of $k$-vectors in the Brillouin zone and whether $N$ and $L$ are even or odd.   \\

\textcolor{black}{The quantity $Z_q$ is a scalar product, $\langle \Psi | \tilde{\Psi}\rangle$, where $|\tilde{\Psi}\rangle = \exp\left(i \frac{2\pi q}{L} \hat{X}\right)|\Psi\rangle$ denotes the ground state  with all momenta shifted by $\frac{2 \pi }{L} q$ as a result of the twist operator.  If the ground state is non-degenerate, then there is only one ground state, which has to have zero total momentum.  Since $Z_q$ will be the scalar product of a zero momentum state and one with a finite momentum (due to the action of $\hat{U}^q$), all $Z_q=0$, except if $q=0$.  When the ground state is degenerate, the ground state wave function will have two components, one with total momentum $\pi/L$, the other with total momentum $-\pi/L$.  Let us write this ground state wave function as
\begin{equation}
|\Psi \rangle = a |\Psi_{\pi/L} \rangle + b |\Psi_{-\pi/L} \rangle,
\end{equation}
where $a$ and $b$ are two complex numbers, each with magnitude $1/\sqrt{2}$, because the total wave function has to have zero total momentum.  To calculate $Z_1$,  we apply the shift operator $\hat{U}$ once, resulting in
\begin{equation}
|\tilde{\Psi} \rangle = \hat{U} |\Psi \rangle =  a |\Psi_{3\pi/L} \rangle + b |\Psi_{\pi/L} \rangle.
\end{equation}
Evaluating the scalar product results in
\begin{equation}
Z_1 = \langle \Psi | \tilde{\Psi} \rangle = a^*b,
\end{equation}
from which it follows that $|Z_1| = 1/2$.  A similar analysis shows that other $Z_q=0$, except $Z_0=1$.  }

The momentum density~\cite{Fazekas99} for both the degenerate and nondegenerate cases is shown in Fig. \ref{fig:nk} in the extended phase indicating discontinuities.   In the degenerate case, states are found with momentum density precisely at the discontinuity, indicated by arrows in the Figure.  The Fourier transform of $Z_q$ gives the polarization distribution (Eq. \ref{eqn:Zq_FT}).  For a metallic system with a nondegenerate ground state this distribution is flat, while for a degenerate system it is RCD (inset of Fig. \ref{fig:U4raisedcos}).  For both functions the GBC (or EK) takes well-known values~\cite{Rinne10} ($U_4=0.4$ and $U_4=0.19792....$, respectively).  Our results based on Eqs. (\ref{eqn:C2C4}) and (\ref{eqn:U4}) do not coincide with those values, but if higher order approximations are used, our $U_4$s converge to the known values.  For the nondegenerate case (flat distribution), this is shown in Fig. 1 of Ref \cite{Hetenyi22}, for the degenerate case (RCD) we present it in Fig. \ref{fig:U4raisedcos} of this work. \\

\textcolor{black}{The results of the above calculations suggest several ways to get around the degeneracy issue.  One is to handle the two cases (degenerate and non-degenerate) separately, since in the former(latter), the GBC will tend to $1/3$($1/2$) in the extended phase.  It is also possible to apply a Peierls phase of $\pi/L$ to lift the degeneracy (or to generate a degenerate ground state).  Such calculations are presented in Figs. \ref{fig:M2} and the main figures of \ref{fig:U4}.  It is also possible to construct a $U_4$ which is not sensitive to this degeneracy.  To do this, send $q \rightarrow 2q$ in all the $Z_q$'s occurring Eq. (\ref{eqn:C2C4}), amounting to doubling the distance over which the finite difference derivative is defined.  Results for a calculation of this type are shown in the inset of Fig. \ref{fig:U4}.  As presented in the Results section, all of these methods work in locating the critical point.}\\

For locating critical points, our results for AA below, and previous studies~\cite{Hetenyi19,Hetenyi22} show that  $U_4$ approximated up to $\mathcal{O}(L^{-2})$ works well.  The original RS variance~\cite{Resta98,Resta99} is based on a finite difference logarithmic derivative and it is correct up to  $\mathcal{O}(L^{-2})$.  It is possible to construct higher order cumulants to any order approximation by extending the RS scheme as well (see, for example, Eq. 51 of Ref. \cite{Hetenyi22}), however, these will fail to give the known values of the flat or the RCD for the EK, because of terms like  $\ln Z_q$ with $q>0$ which diverge as  $Z_q \rightarrow 0$.  \\ 

\section{Results}
\label{sec:results}

\begin{figure}[ht]
 \centering
 \includegraphics[width=8.5cm,keepaspectratio=true]{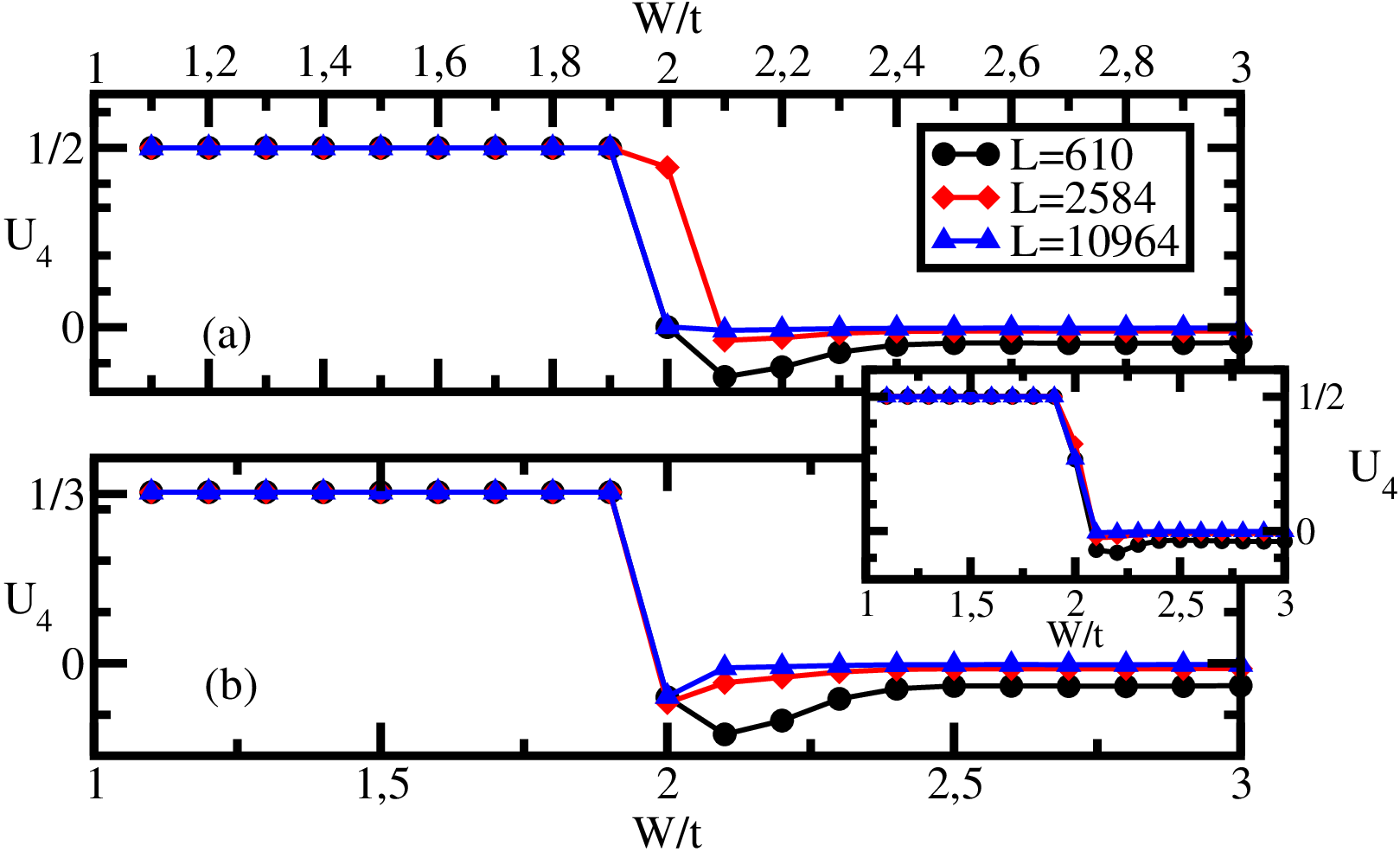}
 \caption{(a) $U_4$ (geometric Binder cumulant) as a function of $W/t$ for system sizes $L=610, 2584, 10964$ for the case with a nondegenerate ground state.  (b) same as (a) but for the case with a degenerate ground state.  The inset shows the same systems as panel (b) but with $U_4$ calculated via a two grid-point approximation.}
 \label{fig:U4}
\end{figure}

Fig. \ref{fig:M2} shows the second cumulant of the polarization as a function of $W/t$ for both nondegenerate (part (a)) and degenerate (part (b)) ground states.   Three system sizes, $L=610, 2584, 10946$, were investigated.     Two sets of results are shown, the upper(lower) panel for a nondegenerate(degenerate) ground state.   In the region $W/t<2$ the functions are straight lines, a drastic change occurs at $W/t=2$, the transition point.  
Defining the size scaling exponent $\gamma$ as $M_2 = a L^\gamma$, we find that for $W/t<2$, $\gamma=2$, while for $W/t>2$, $\gamma=1$ (negligible error in both cases).   These results are in line with expectations~\cite{Chiappe18}.    At the transition point itself, we find  $\gamma=1.44(7)$($\gamma=1.71(3)$) for the nondegenerate(degenerate) case.  The phase transition point is clearly identified.  For comparison, Ref. \cite{Varma15}  reports a flat variance and  $\gamma = 2.008(5)$ for OBC, meaning that our method for PBC, in this sense, coincides with their OBC results.  \\

 The Binder cumulant results (Fig. \ref{fig:U4}) corroborate the findings based on the variance.  In the regime of extended states the Binder cumulant shows the predicted sensitivity to ground state degeneracy, because its value is $1/2$ ($1/3$) for the nondegenerate(degenerate) case.  In the localized regime the GBC becomes negative, but tends to zero with both increasing disorder strength and system size.    \\

\begin{figure}[ht]
 \centering
 \includegraphics[width=8cm,keepaspectratio=true]{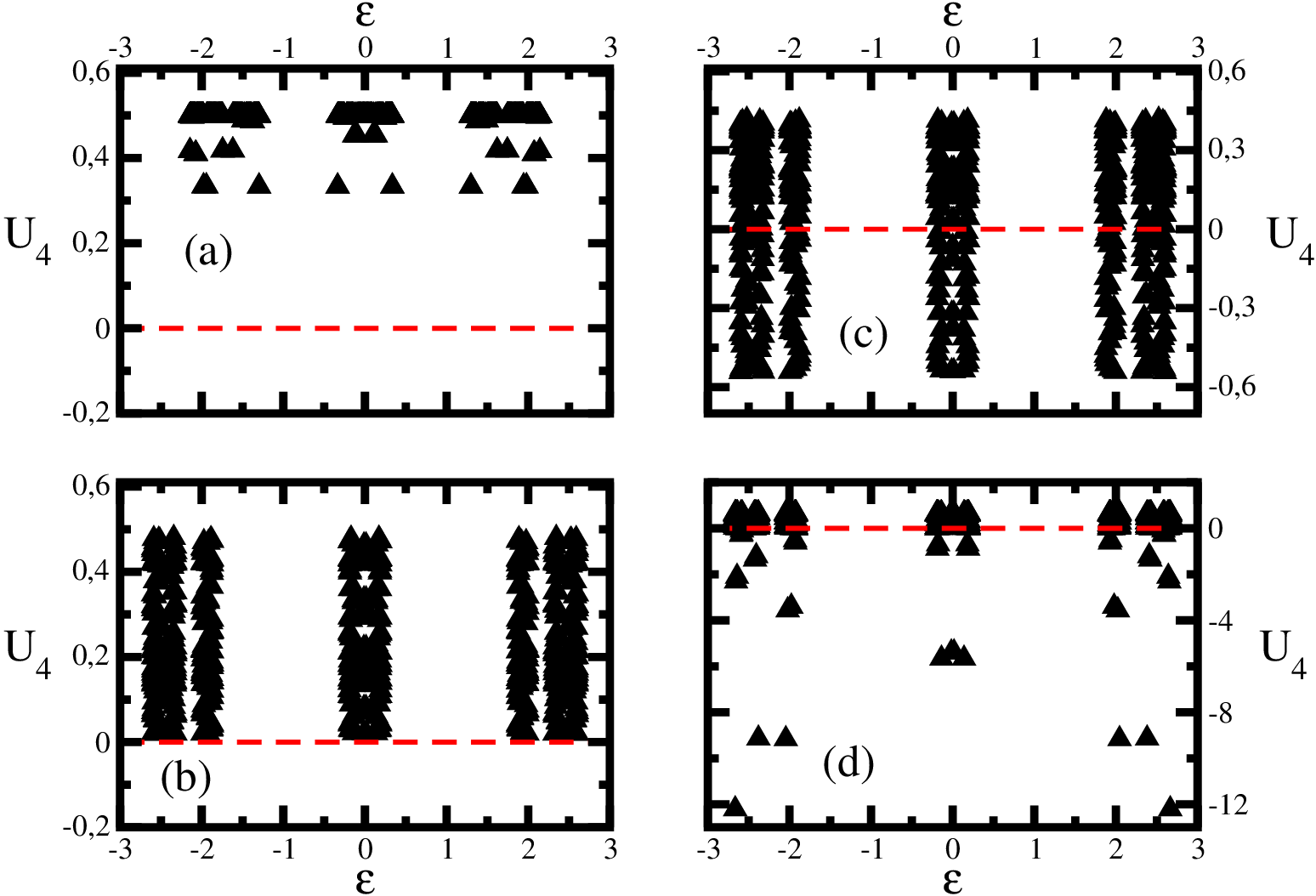}
 \caption{Geometric Binder cumulant $U_4$ of a a given energy eigenstate as a function of energy eigenvalue of the Aubry-Andr\'{e} model for different values of $W/t$: (a) $W/t=1.00$, (b) $W/t=1.99$, (c) $W/t=2.01$, and (d) $W/t=2.10$.   The red dashed line indicates $U_4=0$.}
 \label{fig:U4_state}
\end{figure}

Fig. \ref{fig:U4} shows that the transition can be determined using one system size only when $U_4$ is used, since even for one system size $U_4$ changes sign.  We now exploit this feature to calculate localization in individual eigenstates of the AA model.   To gain insight into the localization mechanism, we calculated $U_4$ for each eigenstate for different values of $W/t$ (Fig. \ref{fig:U4_state}).   For $W/t=0$ (not shown) all states are extended ($U_4=\frac{1}{2}$).  The states are evenly distributed, no clustering occurs, there are no gaps.  As $W$ is increased, but still metallic, bands form, with gaps between them, clustering is seen in panels (a) and (b) of Fig. \ref{fig:U4_state}.  While the values of $U_4$ span a wider range, all states have $U_4>0$.  As the phase boundary is crossed, a fraction of states localize, $U_4<0$.  Interestingly, not all states have $U_4<0$, some appear to remain delocalized, at least in the periodic box of size $L=610$.    Unlike in uncorrelated disordered systems~\cite{Abrahams79} with mobility edges, the localized states are not necessarily on the edges of the bands.   In the AA in 1D, at half filling, the wavefunction is a Slater determinant of occupied states, the localized states localize the whole system.  Away from half-filling the critical point remains $W/t=2$.  This follows from the results in Fig. \ref{fig:U4}.   \\

\section{Conclusion}
\label{sec:conclusion}

We developed a finite size scaling method to be used to locate quantum phase transition points in the context of the modern polarization theory, where the polarization is expressed as a geometric phase, rather than the expectation value of an operator.  The method was applied to the Aubry-Andr\'{e} model, the canonical  model in the study of quasi-periodicity, and one which exhibits a transition between extended and localized phases.  The Binder cumulant construction allows the determination of localization of individual states, because, unlike when the variance is used, a comparison between different system sizes is not necessary.  We calculated the localization of individual eigenstates and found that some of the states remain delocalized even upon crossing the phase transition point.   This result raises interesting questions about localization, and how it occurs~\cite{Jitomirskaya99,Wang17,Zhang15} when periodic boundary conditions are used.

\textcolor{black}{In a seminal paper in 1964, Walter Kohn~\cite{Kohn64} was the first to point out that the quantum criterion to distinguish conductors from insulators is the localization of the center of mass of the charge distribution, rather than the localization of individual charge carriers, which is the appropriate criterion in the classical case.  The numerical testing of this idea was not possible at the time, because in model calculations periodic boundary conditions are used, and the position operator is ill-defined.  The modern theory of polarization, developed~\cite{King-Smith93,Resta94} in the 1990s, overcame this problem by casting the polarization as a geometric phase~\cite{Berry84,Zak89}.   All example calculations based on this theory support the original tenet of Kohn.  Our work provides for the use of scaling methods~\cite{Fisher72a,Fisher72b,Binder81a,Binder81b}, originating from renormalization group theory, in a modern polarization theoretical context, and allows for quantitative tests of Kohn's tenet.}

\section*{Acknowledgments}  I thank M. V. Berry for helpful discussions. I was supported by the National Research, Development and Innovation Fund of Hungary within the Quantum Technology National Excellence Program (Project Nr. 2017-1.2.1-NKP-2017-00001) and by the
BME-Nanotechnology FIKP grant (BME FIKP-NAT).  
\\


\begin{thebibliography}{9}
\bibitem{note}  In lattice models, there is a contribution to the polarization from the nuclei, which shifts the average of the polarization of the electrons.  In this work we are considering centered moments and cumulants which are unaffected by this shift.
\end{thebibliography}
\end{document}